\documentclass[superscriptaddress,aps,nofootinbib,preprint,showpacs]{revtex4}
\usepackage{amssymb}
\usepackage{amsmath}
\usepackage{amsfonts}
\usepackage{graphicx}

\begin{document}
\rightline{\textbf{IPPP/02/47}}
\rightline{\textbf{DCPT/02/94}}
\rightline{\textbf{FISIST/15\,-2002/CFIF}}
\vspace{.3cm}

\title{Flavor-dependent $CP$ violation and electroweak
baryogenesis in supersymmetric theories}

\author{D. Del\'{e}pine}
\email{delepine@cfif.ist.utl.pt} \affiliation{Centro de F\'{\i}sica das
Interac\c{c}\~{o}es Fundamentais (CFIF), Departamento de F\'{\i}sica,
Instituto Superior T\'{e}cnico, Av.Rovisco Pais, 1049-001 Lisboa,
Portugal}

\author{R. Gonz\'{a}lez Felipe}
\email{gonzalez@gtae3.ist.utl.pt} \affiliation{Centro de F\'{\i}sica das
Interac\c{c}\~{o}es Fundamentais (CFIF), Departamento de F\'{\i}sica,
Instituto Superior T\'{e}cnico, Av.Rovisco Pais, 1049-001 Lisboa,
Portugal}

\author{S. Khalil}
\email{shaaban.khalil@durham.ac.uk} \affiliation{IPPP, Physics Department,
Durham University, DH1 3LE, Durham, UK}
\affiliation{Ain Shams University, Faculty of Science, Cairo, 11566,
Egypt}

\author{A.M. Teixeira}
\email{ana@cfif.ist.utl.pt} \affiliation{Centro de F\'{\i}sica das
Interac\c{c}\~{o}es Fundamentais (CFIF), Departamento de F\'{\i}sica,
Instituto Superior T\'{e}cnico, Av.Rovisco Pais, 1049-001 Lisboa,
Portugal}

\begin{abstract}
We analyze electroweak baryogenesis in supersymmetric theories with
flavor-dependent $CP$-violating phases. We generalize the standard
approach to include the flavor dependence of the $CP$-violating sources
and obtain an analytical approximate expression for the baryon asymmetry
of the universe induced by these sources. It is shown that in the
framework where the $\mu$-term is real and the chargino sources vanish,
large flavor mixing might lead to a substantial baryon asymmetry through
the squark contributions, once the condition to have a strong first-order
phase transition induced by light right-handed up squarks is relaxed. We
derive model independent bounds on the relevant up-squark left-right mass
insertions. We show that in supersymmetric models with non-minimal flavor
structure these bounds can be reached and the required baryon asymmetry
can be generated, while satisfying the constraints coming from the
electric dipole moments.

\end{abstract}

\maketitle

\section{Introduction}
\label{intro}

During the last few years, the data collected from the acoustic peaks in
the cosmic microwave background radiation~\cite{Jungman:1995bz} has
allowed to obtain a more precise measurement of the baryon asymmetry of
the universe (BAU). This is expected to further improve in the near future
with the MAP experiment~\cite{MAP} and the PLANCK satellite~\cite{PLANCK}.
At the present time, the measurement of the baryon-to-entropy ratio is
\begin{equation} \label{nBsbound}
0.7 \times 10^{-10} \lesssim \frac{n_B}{s} \lesssim 1.0 \times 10^{-10}\ ,
\end{equation}
where $s=2\pi^2 g_\ast T^3/45 $ is the entropy density and $g_\ast$ is the
effective number of relativistic degrees of freedom.

It is well known that in order to obtain an asymmetry starting from a
symmetric state with a vanishing baryon number, three requirements must be
satisfied: baryon number violation, $C$ and $CP$ violation, and departure
from thermal and kinetic equilibrium~\cite{Sakharov:dj}. In the standard
model (SM) of electroweak interactions, the main source of $CP$ violation
comes from the phase $\delta_{CKM}$ in the Cabibbo-Kobayashi-Maskawa (CKM)
quark mixing matrix. Although this phase is able to account for the
experimentally observed $CP$ violation in the neutral $K$-mesons and, as
recently observed, in the $B_d$ system, it has been shown that it is not
possible to generate sufficient BAU through
$\delta_{CKM}$~\cite{Farrar:sp}. Furthermore, the strength of the phase
transition is too weak in the SM and the universe is approximately in
equilibrium \cite{Buchmuller:1994qy} (for reviews on electroweak
baryogenesis, see for instance Refs.~\cite{Cohen:1993nk,Riotto:1999yt}).

In the context of supersymmetric (SUSY) extensions of the SM, it has been
pointed out that the above problems can be in principle overcome
\cite{Giudice:hh}. Moreover, in the presence of light stops the
electroweak phase transition can be strong enough for baryogenesis to take
place~\cite{Carena:1996wj,
Espinosa:1996qw,Delepine:1996vn,Cline:1996cr,Laine:1996ms,Laine:1998qk,Quiros:2000wk}.
Moreover, SUSY models contain new $CP$-violating sources beyond
$\delta_{CKM}$, namely the Higgs bilinear term, $\mu$, and the soft
breaking terms (gaugino and squark soft masses, bilinear and trilinear
couplings). These can be classified as flavor-blind or flavor-dependent.
The first category includes the phases of the $\mu$- and $B$-parameters,
of the gaugino masses and the overall phase of the trilinear couplings
$A_{ij}$. Two of these phases can be eliminated by $U(1)_{R}$ and
$U(1)_{PQ}$ transformations. The second category contains the phases of
$A_{ij}$ (after the overall phase is factored out), as well as the ones
appearing in the off-diagonal elements of the soft squark masses. The low
energy implications of these flavor-dependent phases on $K$ and $B$ meson
$CP$-violating observables, as well as in rare decays, have been
extensively studied in the
literature~\cite{Khalil:1999zn,Barbieri:1999ax,Branco:2002kz}. The
question that naturally arises is whether the new SUSY phases could
significantly enhance the $CP$-violating sources, so that supersymmetric
electroweak baryogenesis could account for the observed baryon asymmetry
of the universe.

A considerable amount of work has been done concerning the implications of
the new flavor-independent $CP$-violating phases on generating an
acceptable value of the baryon
asymmetry~\cite{Huet:1995sh,Riotto:1997vy,Carena:1997gx,Cline:2000kb,Kainulainen:2001cn,Carena:2002ss}.
It has been shown that in the minimal supersymmetric extension of the SM
(MSSM), if the relative phase $\phi_{\mu}$ between the gaugino soft mass
and the $\mu$ term is not too small, $\phi_{\mu} \gtrsim 0.04$, a
considerable BAU can be generated through the scattering of the charginos
with the bubble wall. However, the non-observation of the electric dipole
moment (EDM) of the electron, neutron and mercury
atom~\cite{Harris:jx,Hagiwara:pw} imposes severe constraints on the
flavor-diagonal phases~\cite{Pokorski:1999hz,Abel:2001vy}, forcing them to
be small. Since in this limit the theory does not acquire any new
symmetry, one has to deal with a naturalness problem, and this is
precisely the so-called SUSY $CP$ problem.

In particular, the EDM's bound the phase $\phi_{\mu}$ to be $\lesssim
10^{-2}$, if the SUSY particle masses are not too heavy ($\lesssim
1$~TeV). It has recently been claimed that new contributions to the EDM of
the electron could eventually rule out the electroweak baryogenesis
scenario based on flavor-independent $CP$-violating
phases~\cite{Chang:1998uc}. A possible way to generate enough BAU while
evading these constraints is to work in the heavy squark limit ($m_Q \sim
3\; \text{TeV}$)~\cite{Pilaftsis:2002fe}.

However, if we assume that SUSY $CP$ violation has a flavor character such
as in the SM~\cite{Abel:2000hn}, one is led to a scenario where all flavor
conserving parameters as the $\mu$-term and gaugino masses are real. In
this framework, the dominant sources of $CP$ violation that are relevant
to electroweak baryogenesis are in general associated with the lightest of
the right-handed up-squarks~\cite{Worah:1997hk}. In the usual MSSM
scenario with complex $\mu$ and gaugino masses, this contribution would be
always subdominant, and could even be neglected when compared to that of
the charginos and neutralinos~\cite{Riotto:1997vy,Carena:1997gx}.

In this paper we study the effects of the flavor-dependent $CP$-violating
phases on the mechanism of electroweak baryogenesis. We show that in
generic SUSY models with non-universal soft SUSY breaking terms, and in
particular with non-universal $A$-terms, the squark contributions to the
BAU are far from being negligible. Although in this framework the
$\mu$-term is real, flavor mixing might lead to a potentially large baryon
asymmetry. As an example,  we analyze the baryon asymmetry in
supersymmetric models with Hermitian flavor structures, a type of model
that also provides an elegant solution for the EDM's
suppression~\cite{Abel:2000hn,Chen:2001cv,Khalil:2002jq}.

Obtaining sufficiently accurate transport equations for particles
propagating in the presence of a $CP$-violating bubble wall at the
electroweak phase transition is crucial for the computation of the baryon
asymmetry generated in the context of electroweak baryogenesis. In spite
of the general consensus on the existence and nature of the $CP$-violating
sources responsible for the baryon production, there is still a
controversy in the literature in what concerns the strength and form of
these sources as well as the transport equations to be used in the
calculations~\cite{Huet:1995sh,Riotto:1997vy,Carena:1997gx,Cline:2000kb,Kainulainen:2001cn,Carena:2002ss}.
For instance, the results obtained in the MSSM framework for the chargino
and squark sources by making use of the continuity equations and the
relaxation time approximation~\cite{Riotto:1997vy,Carena:1997gx} are
different from those based on WKB-methods~\cite{Cline:2000kb}.
Furthermore, it has been claimed in
Ref.~\cite{Cline:2000kb,Kainulainen:2001cn} that the dominant contribution
to the chargino source,  typically found within the MSSM and which is of
the form $\epsilon_{ij} v_i \partial_\mu v_j$ ($v_{1,2}$ are the
expectation values of the two Higgs doublets)
\cite{Riotto:1997vy,Carena:1997gx}, is absent, thus leading to a
suppressed baryon asymmetry \cite{Cline:2000kb,Kainulainen:2001cn}.
However, in a recent work~\cite{Carena:2002ss} it is argued that such a
suppression is in fact an artifact of the approximation used by the
authors of Refs.~\cite{Cline:2000kb,Kainulainen:2001cn} in order to
compute the $CP$-violating currents. In view of the above discussion, we
adopt in the present work the approach of
Refs.~\cite{Huet:1995sh,Riotto:1997vy,Carena:1997gx,Carena:2002ss} in
order to derive the $CP$-violating currents and sources.

The paper is organized as follows. By using the closed time path
formalism, we build in Section~\ref{sources} the quantum Boltzmann
equations to obtain the $CP$-violating sources relevant to baryogenesis
and discuss their general flavor dependence. In Section~\ref{ebau} we
compute the baryon asymmetry induced by these $CP$-violating sources. An
analytical approximate expression for the asymmetry is also presented. In
Section~\ref{textures} we obtain model independent bounds for the
up-squark left-right mass insertions by requiring the baryon-to-entropy
ratio to lie in the experimental range. We also consider some specific
SUSY models with minimal and non-minimal flavor structure and discuss
whether or not they can account for the required BAU. Our numerical
results are presented in Section~\ref{numerical}. Finally, we present our
concluding remarks in Section~\ref{conclusion}.

\section{$CP$-violating sources for the baryon asymmetry}\label{sources}

Before computing the $CP$-violating sources for the baryon asymmetry, we
shall briefly recall some characteristics of non-equilibrium quantum field
theory \cite{Kadanoff:1962,Chou:es}. During a first-order phase
transition, the thermodynamical system is far from equilibrium. In order
to keep an explicit time dependence, one should use the real-time finite
temperature quantum field theory. The most used and powerful formalism to
describe such a system is the so-called closed time path formalism (CTP),
which is a generalization of the time contour integration to a closed time
path. More precisely, the time integration is deformed to run from
$-\infty$ to $+\infty$ and back to $-\infty$. The main effect of this
closed time path is to double the field variables so that for each field
we have four different real-time propagators on the contour. In the case
of a boson field $\phi$, we can write the corresponding Green functions in
terms of the $2 \times 2$ matrices \cite{craig1968}
\begin{equation}
\label{2:mxGtilde} \widetilde{G}_{\phi}=\left(
\begin{array}[c]{cc}
G_{\phi}^{t} & \;\;G_{\phi}^{<}\\
G_{\phi}^{>} & - G_{\phi}^{\overline{t}}
\end{array}
\right)\ ,
\end{equation}
where
\begin{align}
G_{\phi}^{>}(x,y) =-i\left\langle \phi(x)\phi^{\dagger}(y)\right\rangle\ ,
\quad G_{\phi}^{<}(x,y) =- i\left\langle
\phi^{\dagger}(y)\phi(x)\right\rangle \label{2:G<>}\ ,
\end{align}
\begin{align}
G_{\phi}^{t}(x,y) &  =\theta(x,y)\;G_{\phi}^{>}(x,y)+\theta(y,x)\;
G_{\phi}^{<}(x,y)\ , \nonumber\\
G_{\phi}^{\bar{t}}(x,y) &  =\theta(y,x)\;G_{\phi}^{>}(x,y)+\theta
(x,y)\;G_{\phi}^{<}(x,y)\label{2:Gt}\;.
\end{align}
In what follows the subscript $\phi$ will be omitted to simplify our
notation.

As mentioned in the introduction, an accurate computation of the
$CP$-violating sources responsible for electroweak baryogenesis is crucial
to obtain a reliable estimate of the baryon asymmetry. Here we adopt the
method developed by Kadanoff and Baym \cite{Kadanoff:1962} to derive the
quantum Boltzmann equations for a generic bosonic particle asymmetry. We
shall compute the sources using the formalism based on CTP and the Dyson
equations as described in Ref.~\cite{Riotto:1997vy}.

In general it is possible to write the Dyson equation for any propagator
$\widetilde{G}$ as
\begin{align}\label{2:Dyson1}
\tilde{G}(x,y) &  =\tilde{G}^{0}(x,y)+\int d^{4} x_{1} \int d^{4}x_{2}\;
\tilde{G}^{0}(x,x_{1})\;
\tilde{\Sigma}(x_{1},x_{2})\;\tilde{G}(x_{2},y)\nonumber\\
&=\tilde{G}^{0}(x,y)+\int d^{4}x_{1} \int d^{4}x_{2} \;
\tilde{G}(x,x_{1})\;\tilde{\Sigma}(x_{1},x_{2})\;
\tilde{G}^{0}(x_{2},y)\;,
\end{align}
where $\tilde{G}^{0}$ are the non-interacting Green functions and
$\tilde{\Sigma}$ are the self-energy functions, which can be also
expressed in terms of the $2 \times 2$ matrices
\begin{equation}
\tilde{\Sigma}=\left(
\begin{array}
[c]{cc}
\Sigma^{t} & \Sigma^{<}\\
\Sigma^{>} & -\Sigma^{\bar{t}}
\end{array}
\right)  \;,\label{2:Sigma}
\end{equation}
satisfying the relations
\begin{align}
\Sigma^{t}(x,y) &
=\theta(x,y)\;\Sigma^{>}(x,y)+\theta(y,x)\;\Sigma^{<}(x,y)\ ,
\nonumber \label{2:Sigmat}\\
\Sigma^{\bar{t}}(x,y) &  =\theta(y,x)\;\Sigma^{>}(x,y)+\theta
(x,y)\;\Sigma^{<}(x,y)\;.
\end{align}

Iterating Eq.~(\ref{2:Dyson1}) one finds
\begin{align}
\label{2:Dyson_iter} \tilde{G}(x,y) &  = \tilde{G}^{0}(x,y)+\int
d^{4}x_{1} \int d^{4}x_{2}\;
\tilde{G}^{0}(x,x_{1})\;\tilde{\Sigma}(x_{1},x_{2})\;
\tilde{G}^{0}(x_{2},y)\nonumber\\
&  +\int d^{4}x_{1} \int d^{4}x_{2} \int d^{4}x_{3} \int d^{4}x_{4}\;
\tilde{G}^{0}(x,x_{1})\;\tilde{\Sigma}(x_{1},x_{2})\;
\tilde{G}^{0}(x_{2},x_{3})\;\tilde{\Sigma}(x_{3},x_{4})\;
\tilde{G}^{0}(x_{4},y)\nonumber\\
& + \ldots\;.
\end{align}

In order to compute the sources for the squark diffusion equations, we
shall write the quantum Boltzmann equations (QBE) for a generic bosonic
particle asymmetry,
\begin{equation}
\label{2:genQBE} \frac{\partial\;n_{\phi}}{\partial\;T}+\pmb{\nabla} \cdot
\pmb{J}_{\phi }=\mathcal{S}\;,
\end{equation}
where $n_{\phi}$ is the number density of particles minus antiparticles
and $\mathcal{S}$ is the associated $CP$-violating source. More
specifically, we shall derive the QBE for the current
\begin{equation}
\label{2:avJ} \langle J_{\phi}^{\mu}(x)\rangle\equiv
i\;\langle\phi^{\dagger} (x)\;\overleftrightarrow{\partial}_{x}^{\mu}\;
\phi(x)\rangle\equiv\left( n_{\phi}(x),\pmb{J}_{\phi}(x)\right)  \;,
\end{equation}
or in terms of the Green functions,
\begin{equation}
\label{2:limJ} \langle J_{\phi}^{\mu}(x)\rangle =
-\operatornamewithlimits{\lim}_{x \rightarrow y}(\partial_{x}^{\mu}-
\partial_{y}^{\mu})\;G^{<}(x,y)\;.
\end{equation}

It proves convenient to work in the coordinate system of the center of
mass, which is defined by
\begin{equation}\label{2:Xcm}
X=(T,\pmb{R})=\frac{1}{2}(x+y)\;,\quad\bar{x}=(t,\pmb {r})=x-y
\nonumber\;.
\end{equation}
Hence
\begin{equation}
\label{2:Gcm} G^{<}(x,y)=G^{<}(t,\pmb{r},T,\pmb
{R})=-i\langle\phi^{\dagger}(T-\frac{t}{2},\pmb{R}-\frac
{\pmb{r}}{2})\;\phi(T+\frac{t}{2},\pmb{R}+\frac {\pmb{r}}{2})\rangle
\nonumber\ .
\end{equation}

The next step is to find a solution for $G^{<}\ $ when the system is not
in equilibrium. This can be achieved by applying the operator
$(\overrightarrow{\square}+m^{2})$ on both sides of the equivalent
representations of the Schwinger-Dyson equation (\ref{2:Dyson1}):
\begin{align}\label{2:OpDyson}
(\overrightarrow{\square}_{x}+m^{2})\;\tilde{G}(x,y) &
=\delta^{4}(x,y)\;\openone +\int d^{4}x_{1}\;
\tilde{\Sigma}(x,x_{1})\;\tilde{G}(x_{1},y)\;,\\
\tilde{G}(x,y)\;(\overleftarrow{\square}_{y}+m^{2}) &  =
\delta^{4}(x,y)\;\openone+\int d^{4}x_{1}\;
\tilde{G}(x,x_{1})\;\tilde{\Sigma}(x_{1},y)\;,
\end{align}
where $\openone$ is the identity matrix. Since $(\Sigma
G)^{<}=\Sigma^{t}\;G^{<}- \Sigma^{<}\;G^{\bar{t}}$ and
$(G\Sigma)^{<}=G^{t}\;\Sigma^{<}-G^{<}\; \Sigma^{\bar{t}}$, the equations
for the $G^{<}$ component read as
\begin{align}\label{2:OpDyson2}
(\overrightarrow{\square}_{x}+m^{2})\;\tilde{G}^{<}(x,y) &  = \int
d^{4}x_{1}\;[\Sigma^{t}(x,x_{1})\;G^{<}(x_{1},y)
-\Sigma^{<}(x,x_{1})\;G^{\bar{t}}(x_{1},y)]\ ,\\
\tilde{G}^{<}(x,y)\;(\overleftarrow{\square}_{y}+m^{2}) &  = \int
d^{4}x_{1}\;[G^{t}(x,x_{1})\;\Sigma^{<}(x_{1},y)-
G^{<}(x,x_{1})\;\Sigma^{\bar{t}}(x_{1},y)]\;.
\end{align}

The variation of the current is thus given by
\begin{align}\label{2:dJdX}
\frac{\partial\;J_{\phi}^{\mu}(X)}{\partial\;X^{\mu}} &  =-\partial_{\mu}
^{X}\left\{\operatornamewithlimits{\lim}_{x \rightarrow
y}(\partial_{x}^{\mu}-\partial_{y}^{\mu})\;G^{<}(x,y)\right\}
\nonumber\\
&  =-\int d^{4}x_{1}\;[\Sigma^{t}(X,x_{1})\;G^{<}(x_{1}
,X)-\Sigma^{<}(X,x_{1})\;G^{\bar{t}}(x_{1},X)\nonumber\\
&  -G^{t}(X,x_{1})\;\Sigma^{<}(x_{1},X)+G^{<}%
(X,x_{1})\;\Sigma^{\bar{t}}(x_{1},X)]\;.
\end{align}

Using Eqs.~(\ref{2:Gt}) and (\ref{2:Sigmat}), we can compute the
$CP$-violating source for a generic squark current:
\begin{align}\label{2:dndT}
\mathcal{S}= \frac{\partial\;n_{\phi}(X)}{\partial\;T}+\pmb{\nabla} \cdot
\pmb{J}_{\phi}(X) & =-\int d^{3}r'
\operatornamewithlimits{\int}_{-\infty}^{\;\;T}dt'\;\left\{
\Sigma^{>}(X,x')\;G^{<}(x',X)-\Sigma^{<}(X,x')G^{>}(x',X)\right. \nonumber\\
&  -G^{>}(X,x')\;\Sigma^{<}(x',X)+\left.
G^{<}(X,x')\;\Sigma^{>}(x',X)\right\} \;.
\end{align}

Inserting the interactions of the $\phi$ field with the background, the
self-energy functions $\Sigma^{>}(x,y)$ can be written as follows
\begin{align}\label{2:Sigma>}
\Sigma^{>}(x,y)  &  =g(x)\;\delta^{4}(x-y)+g (x)\; {G^{0}}^{>}(x,y)\;g(y)
+ \int d^{4}z\;\left[ g(x)\;{\tilde{G^{0}}}(x,z)\;g
(z)\;{\tilde{G^{0}}}(z,y)\;g(y)\right]^{>}\nonumber\\
&  +  \int d^{4}w\;\int d^{4}z\; \left[ g(x)\;{\tilde{G^{0}}
}(x,w)\;g(w)\;{\tilde{G^{0}}}(w,z)\; g(z)\;{\tilde{G^{0}}
}(z,y)\;g(y)\right]^{>}+\ldots\ ,
\end{align}
where the scalar function $g(x)$ describes the interaction of the field
$\phi$ at position $x$ with the background fields.

\begin{figure}[ht]
\begin{center}
\includegraphics[width=9cm]{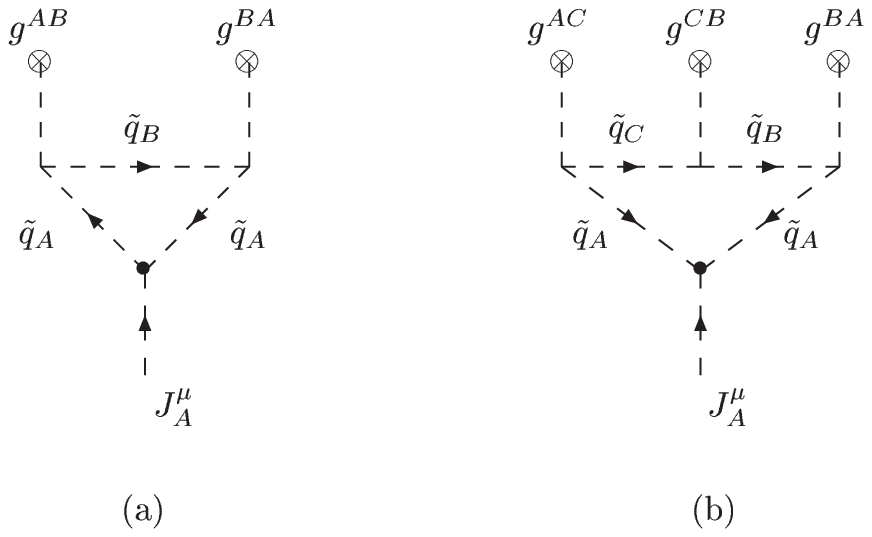} \\
\caption{A symbolic representation of the structure of the $CP$-violating
squark source.} \label{fig1}
\end{center}
\end{figure}

Taking into account the chiral and flavor structures of the squarks, this
generic squark source can be written as
\begin{align}\label{2:SijAB}
\mathcal{S}_{ij}^{AB} = & -\int d^{3}r'
\operatornamewithlimits{\int}_{-\infty}^{\;\;T}dt'\; \left\{
\left[g(X)\right] _{ij}^{AB}\;\delta^{4}(X-x')\;
\delta_{ij}^{AB}\;[{G}^{<}(x',X)]_{j}^{B}
 \right. \nonumber\\
& -\left[ g(X)\right]_{ij}^{AB}\;\delta^{4}(X-x')\;
\delta_{ij}^{AB}\;[{G}^{>}(x',X)]_{j}^{B}\nonumber\\
& -[{G}^{>}(X,x')]_{i}^{A}\;\left[g(x')\right]_{ij}^{AB}\;
\delta^{4}(X-x')\;\delta_{ij}^{AB} \nonumber\\
& +[{G}^{<}(X,x')]_{i}^{A}\; \left[g_(x')\right]_{ij}^{AB}\;
\delta^{4}(X-x')\;\delta_{ij}^{AB}\nonumber\\
& +\left[g(X)\right]_{ik}^{AC}\;[{G}^{>}(X,x')]_{k}^{C}\;
\left[g(x')\right]_{kj}^{CB}\;[{G}^{<}(x',X)]_{j}^{B}\nonumber\\
&-\left[g(X)\right]_{ik}^{AC}\;[{G}^{<}(X,x')]_{k}^{C}\;
\left[g(x')\right]_{kj}^{CB}\;[{G}^{>}(x',X)]_{j}^{B}
\nonumber\\
&-[{G}^{>}(X,x')]_{i}^{A}\;\left[g(x')\right]_{ik}^{AC}\;
[{G}^{<}(x',X)]_{k}^{C}\;\left[g(X)\right]_{kj}^{CB}\nonumber\\
& \left. + [{G}^{<}(X,x')]_{i}^{A}\;
\left[g(x')\right]_{ik}^{AC}\;[{G}^{>}(x',X)]_{k}^{C}\;
\left[g(X)\right]_{kj}^{CB} + \mathcal{O}([g(x)]^3) \right\} \;,
\end{align}
where $(i,j,k)$ are flavor indices and $(A,B,C)$ refer to $L,R$
chiralities. In the above expression, we have kept only terms up to second
order in $g(x)$. Note that this formula is valid for squarks of any flavor
and/or chirality. Thus, we have
\begin{equation}\label{2:deltaAB}
\partial_{\mu}^{X}\;J_{\phi_{i}^{A}}^{\mu}\;=\;\delta_{AB} \;\delta^{ij}\;
\mathcal{S}_{ij}^{AB}= \mathcal{S}_{ii}^{AA}\;,
\end{equation}
with $\phi_i^L \equiv \tilde{U}_{Li}$ and $\phi_i^R \equiv
\tilde{U}_{Ri}$, the left and right-handed up-squark fields, respectively.
As expected, the contributions of the terms proportional to
$\delta^{4}(X-x')$ vanish\footnote{Throughout we are using the relations:
$G^{<}(x)=G^{>}(-x)\;, \quad {[G^{<}(x)]}^{*}=-G^{<}(-x)\;, \quad
{[G^{>}(x)]}^{*}=-G^{>}(-x)\;,$ which imply $G^{<}(x)={[G^{>}(x)]}^{*}$.}.
The structure of the source can be symbolically depicted as in
Fig.~\ref{fig1}, where the $\otimes$ denotes interactions with the
background fields, parametrized by the scalar function $g(x)$.

Before proceeding, let us consider the Higgs-squark interaction terms,
arising from both SUSY conserving and SUSY breaking terms, that can be
parametrized in the following way (flavor indices are omitted) :
\begin{equation}\label{2:Lgx}
\mathcal{L}_{\tilde{q}}= \tilde{U}_L^* \left[g(X)\right]^{LL} \tilde{U}_L+
\tilde{U}_R^* \left[g(X)\right]^{RR} \tilde{U}_R+ (\tilde{U}_L^*
\left[g(X)\right]^{LR} \tilde{U}_R+ \text{H.c})\ .
\end{equation}
Note that in addition to the interactions with the background, the
functions $\left[g(X)\right]^{AB}$ contain flavor and chirality mixing
terms that originate from the soft breaking SUSY Lagrangian and become
relevant when one computes higher order terms like those appearing in
Fig.~\ref{fig1}(b).

Within a generic supersymmetric extension of the SM, the functions $g(X)$
can be parametrized as follows:
\begin{align}\label{2:gXAB}
\left[{g}(X)\right]_{ij}^{LL}  & =\left(M_{\tilde{Q}}^{2}+
v_{2}^{2}(X)\;h_{u}^{*}\;h_{u}^{T}+D_L^2(X)\right)_{ij}\;,\nonumber\\
\left[{g}(X)\right]_{ij}^{RR}  & =\left(M_{\tilde{U}_{R}}^{2}+
v_{2}^{2}(X)\;h_{u}^{T}\;h_{u}^{*}+D_R^2(X)\right)_{ij}\;,\nonumber\\
\left[{g}(X)\right]_{ij}^{LR}  & =\left(-v_{1}(X)\;\mu\;h_{u}^{*}+
v_{2}(X)\;{(Y_{u}^{A})}^{*}\right)_{ij}\;,\nonumber\\
\left[{g}(X)\right]^{RL}  & ={\left[{g}(X)\right]^{LR}}^{\dagger}\;.
\end{align}
In the above equations, $h_u$ denotes the Yukawa coupling matrix for the
up quarks and $v_{1,2}(x)$ are the $x$-dependent vacuum expectation values
of the MSSM Higgs fields $H_{1,2}\ $, defined as $v_{1}= v \cos \beta/
\sqrt 2$ and $ v_{2}= v \sin \beta/ \sqrt 2$, with $\tan \beta = v_2/v_1$.
$M_{\tilde{Q}}$ and $M_{\tilde{U}_{R}}$ are the squark soft breaking mass
matrices, $\mu$ is the Higgs bilinear term and $Y_{u}^{A}$ the soft
trilinear terms, which are decomposed as
\begin{equation}\label{2:YAdef}
(Y_{u}^{A})_{ij} \equiv (h_u)_{ij} \;A^u_{ij}\;,
\end{equation}
with no summation over $i,j$.

The $D$-term contributions are given by
\begin{align}\label{2:DLDR}
D_L^2(X)=&\left(\frac{1}{2} -\frac{2}{3} \sin^2 \theta_W \right)\;
\frac{g^2+{g^\prime}^2}{2}\; (v_{1}^{2}(X)-v_{2}^{2}(X))\;, \nonumber \\
D_R^2(X)=&\left(\frac{2}{3} \sin^2 \theta_W \right)\;
\frac{g^2+{g^\prime}^2}{2}\; (v_{1}^{2}(X)-v_{2}^{2}(X))\;,
\end{align}
$g, g'$ are the $SU(2)$ and $U(1)$ weak couplings, respectively.  These
terms can be neglected at the electroweak phase transition since they are
suppressed by the weak couplings when compared to the terms proportional
to the top Yukawa coupling.

Since we are interested in performing a basis independent calculation, we
should express the current as an invariant quantity. Let us define the
invariant current as
\begin{equation}\label{2:JtrS}
\partial_{\mu}^{X}\;J_{\phi^{A}}^{\mu}\;=
\;\operatornamewithlimits{\sum}_{i=1}^{n_f}\;\mathcal{S}_{ii}^{AA}\;=
\;\operatorname{Tr}\;\mathcal{S}^{AA}\;,
\end{equation}
where $n_{f}$ is the number of up-squark flavors.

Under a generic rotation that transforms the left- and right-handed up
squarks as $ \tilde{U}_{L}\rightarrow W_{L}\;\tilde{U}_{L}\ ,\;
\tilde{U}_{R}\rightarrow Z_{R}\;\tilde{U}_{R}\;,$ we have
\begin{align}\label{2:Gtransf}
& {\left[{G(x,y)}^{>,<}\right]}^{L}\rightarrow W_{L}\;
{\left[{G(x,y)}^{>,<}\right]}^{L}\;W_{L}^{\dagger}
\equiv{\left[{{G}(x,y)}^{>,<}\right]}^{L}\ ,\nonumber\\
& {\left[{G(x,y)}^{>,<}\right]}^{R}\rightarrow Z_{R}\;
{\left[{G(x,y)}^{>,<}\right]}^{R}\;Z_{R}^{\dagger}
\equiv{\left[{\bar{G}(x,y)}^{>,<}\right]}^{R}\;,
\end{align}
\begin{align}
\left[g(x)\right]^{LL}\rightarrow W_{L}\;
\left[g(x)\right]^{LL}\;W_{L}^{\dagger}  &
\equiv\left[\bar{g}(x)\right]^{LL}\ ,\nonumber\\
\left[g(x)\right]^{LR}\rightarrow W_{L}\;
\left[g(x)\right]^{LR}\;Z_{R}^{\dagger}  &
\equiv\left[\bar{g}(x)\right]^{LR}\ ,\nonumber\\
\left[g(x)\right]^{RR}\rightarrow Z_{R}\;
\left[g(x)\right]^{RR}\;Z_{R}^{\dagger}  &
\equiv\left[\bar{g}(x)\right]^{RR}\ .\label{2:gtransf}
\end{align}

Thus the invariant source which is given by
\begin{align}\label{2:trSAA}
\operatorname{Tr}\;\mathcal{S}^{AA}  = &-\int
d^{3}r'\operatornamewithlimits{\int}_{-\infty}^{\;\;T}dt'
\sum_{i,k=1}^{n_{f}}\;\left\{\left[g(X)\right]_{ik}^{AC}\;
[{G}^{>}(X,x')]_{k}^{C}\;\left[g(x')\right]_{ki}^{CA}
\;[{G}^{<}(x',X)]_{i}^{A}\right. \nonumber\\
&  -\left[g(X)\right]_{ik}^{AC}\;\left.
[{G}^{<}(X,x')]_{k}^{C}\;\left[g(x')\right]_{ki}^{CA}
\;[{G}^{>}(x',X)]_{i}^{A}+(X\leftrightarrow x')+\ldots \right\}  ,
\end{align}
can be rewritten in any weak basis. The above equation can be recast in
the form
\begin{align}\label{2:trSAA2}
\operatorname{Tr}\;\mathcal{S}^{AA}  = &-2\;i\; \int d^{3}r'
\operatornamewithlimits{\int}_{-\infty}^{\;\;T}dt'\sum_{i,k=1}^{n_{f}
}\;\left\{  \operatorname{Im}\,\left(  [{\bar{G}}^{<}(X,x')]_{k}
^{C}\;[{\bar{G}}^{<}(x',X)]_{i}^{A}\right)  \right. \nonumber\\
& \times  \left.  \left(  \left[  \bar{g}(X)\right]  _{ik}^{AC}\;\left[
\bar{g}(x')\right]  _{ki}^{CA}-\left[  \bar{g}(x')\right]
_{ik}^{AC}\;\left[ \bar{g}(X)\right]_{ki}^{CA}\right) +\ldots \right\} .
\end{align}

Finally we obtain
\begin{align}\label{2:SRR}
\operatorname{Tr}\;\mathcal{S}^{RR}= &\; 4\int d^{3}r'
\operatornamewithlimits{\int}_{-\infty}^{\;\;T}dt' \left(
v_{1}(X)\;v_{2}(x') - v_{2}(X)\;v_{1}(x')\right) \sum_{i,k=1}^{n_{f}}\;
\left\{\operatorname{Im}\;\left([{\bar{G}}^{<}(X,x')]_{k}^{L}\;
[{\bar{G}}^{<}(x',X)]_{i}^{R}\right) \right. \nonumber\\
& \times \left. \operatorname{Im}\;\left(\mu\left[Z_{R}\;{(Y_{u}^{A})}^{T}
\;W_{L}^{\dagger}\right]_{ik}\;\left[W_{L}\;h_{u}^{\ast}
\;Z_{R}^{\dagger}\right]_{ki}\right) \right\}\;.
\end{align}

A convenient basis for the current associated with the right-handed
up-squark $\tilde{U}_{Ri}$ is that where the Green functions are diagonal.
For simplicity, we shall assume that the left-handed squark mass matrix
$\tilde{M}_{LL}$ is already diagonal. Therefore one has $W_{L}=\openone$.
Moreover, we will also assume that the left-handed squarks are heavy ($m_Q
\gg m_t$) and nearly degenerate, so that $\tilde{M}_{LL} \simeq
\operatorname{diag}(m_Q, m_Q, m_Q)$ and ${\bar{G}}(x)_{k}^{L}\; \equiv
\;{\bar{G}}(x)^{L}$ for any flavor. The right-handed squark mass matrix
$\tilde{M}_{RR}$ is diagonalized by a unitary matrix $Z_R$ so that
\begin{equation}\label{2:WLZR}
Z_{R}\tilde{M}_{RR}^{2} Z_{R}^{\dagger}=d_{RR}^{2}\;,
\end{equation}
$d_{RR} = \operatorname{diag}(m_{R1}, m_{R2}, m_{R3})$.
 In this case,
\begin{align}\label{2:trSRR}
\text{Tr}\;\mathcal{S}^{RR}= &\; 4 \int d^{3}r'
\operatornamewithlimits{\int}_{-\infty}^{\;\;T}dt'
\left(v_{1}(X)\;v_{2}(x')-v_{2}(X)\;v_{1}(x')\right) \sum_{i=1}^{n_{f}}
\left\{\operatorname{Im}
\left([{\bar{G}}^{<}(X,x')]^{L}\;[{\bar{G}}^{<}(x',X)]_{i}^{R}\right)
\right. \nonumber \\
& \times  \left.
{\operatorname{Im}}\left(\mu\left[Z_{R}\;{(Y_{u}^{A})}^{T}\;
h_{u}^{\ast}\;Z_{R}^{\dagger}\right]_{ii}\right) \right\}\;.
\end{align}

In order to obtain an analytical expression, it is useful to perform an
expansion in the bubble wall velocity $v_{w}$. Such an expansion is well
justified in the case of the MSSM, since bubbles are typically formed with
thick walls ($L_w \sim (10-100)/T$) and propagate with extremely
nonrelativistic velocities ($v_{w} \sim 0.1-0.01$)
\cite{Moreno:1998bq,Moore:2000wx}. Using the derivative expansion
\begin{equation}\label{2:expansion_vev}
v_{i}(x)=\sum_{n=0}^{\infty}\frac{1}{n!} \frac{\partial^{n}}{(\partial
X^{\mu})^{n}}\; v_{i}(X) \left(x^{\mu}-X^{\mu}\right)^{n}\;,
\end{equation}
it is easy to check that the first nonzero contribution to
Eq.~(\ref{2:trSRR}) is given by the $n=1$ term in the expansion
(\ref{2:expansion_vev}). The associated contribution to the source is
proportional to the function
\begin{equation}\label{2:betaX}
v_{1}(X)\;\partial_{X}^{\mu}\;v_{2}(X)-v_{2}(X)\;\partial_{X}^{\mu}\;v_{1}(X)\equiv
\frac{1}{2}\;v^{2}(X)\;\partial_{X}^{\mu}\beta(X)\;,
\end{equation}
which in turn implies that the source (\ref{2:trSRR}) is linear in the
wall velocity $v_{w}$. Neglecting terms of higher order in $v_w$, which
amounts to using the thermal equilibrium Green functions, one can obtain
an explicit expression for the dominant contribution:
\begin{equation}\label{2:baryo_eqs}%
\operatorname{Tr}\;\mathcal{S}^{RR}= v^{2}(X)\;
\partial_{X} \beta(X) \; \left\{\sum_{i=1}^{n_{f}}I_{RR}^{i}
\operatorname{Im}\left(\mu\left[Z_{R}\;{(Y_{u}^{A})}^{T}\;h_{u}^{\ast}\;
Z_{R}^{\dagger}\right]_{ii}\right) \right\}\;,
\end{equation}
with $I_{RR}^i$ given by \cite{Riotto:1997vy}
\begin{align}\label{2:IRR}
I_{RR}^{i} & = \operatornamewithlimits{\int}_{0}^{\;\;\infty} dk\;
\frac{k^{2}}{4 \pi^{2}\omega_{Q}\; \omega_{R}^{i}}
\;\left[\left(1+2\;\mathrm{\operatorname{Re}}\;(n_{Q})\right)\;
I(\omega_{R}^{i},\Gamma_{R}^{i},\omega_{Q},\Gamma_{Q})\right. \nonumber\\
&  \left. +\left(1+2\;\mathrm{\operatorname{Re}}\;(n_{R}^{i})\right)\;
I(\omega_{Q},\Gamma_{Q},\omega_{R}^{i},\Gamma_{R}^{i})
+2\left(\mathrm{\operatorname{Im}}\;(n_{R}^{i})+
\mathrm{\operatorname{Im}}\;(n_{Q})\right)\;
G(\omega_{R}^{i},\Gamma_{R}^{i},\omega_{Q},\Gamma_{Q})\right]\;.
\end{align}
The squark equilibrium distribution functions are
\begin{equation}\label{2:BEeqdistr}
n_{R,Q}^{i} = 1/\left[\exp \left( \omega_{R,Q}^{i}\;/\;T+
i\Gamma_{R,Q}^{i}\;/\;T \right)-1\right]\;,
\end{equation}
where the finite widths $\Gamma_{R,Q}^{i}$ account for the interactions
with the plasma,
\begin{align}\label{2:omegaLR}
\omega_{Q}^{2} = \pmb{k}^{2}+ m_{Q}^{2}+ \bar{m}_{Q}^{2}(\phi,T)\;, \quad
\omega_{Ri}^{2} = \pmb{k}^{2}+m_{Ri}^{2}+ \bar{m}_{Ri}^{2}(\phi,T)\ ,
\end{align}
and $\bar{m}_{Q,Ri}^{2}(\phi,T)$ are the field-dependent contributions to
the squark masses, which include the temperature dependent self-energies
$\Pi_{Q,Ri}(T)$ \cite{Comelli:1996vm}. The functions $I$ and $G$ can be
written as follows:
\begin{align}\label{2:IGfunctions}
I(a,b,c,d)  = I_{+}(a,b,c,d)+I_{-}(a,b,c,d)\ , \quad G(a,b,c,d)  =
G_{+}(a,b,c,d)+G_{-}(a,b,c,d)\ ,
\end{align}
\begin{align} \label{2:IpGp}
I_{\pm}(a,b,c,d)  & = \frac{1}{2}\frac{1}{(a \pm c)^{2}+(b+d)^{2}} \;\sin
\left( 2 \arctan \frac{a \pm c}{b+d}\right)\ ,\nonumber\\
G_{\pm}(a,b,c,d)  & = \frac{1}{2}\frac{1}{(a \pm c)^{2}+(b+d)^{2}} \;\cos
\left( 2 \arctan \frac{a \pm c}{b+d}\right)\ .
\end{align}

Before proceeding to the explicit computation of the baryon asymmetry, it
is worth emphasizing that from Eqs.~(\ref{2:baryo_eqs}) and (\ref{2:IRR}),
the dominant contribution to $\text{Tr}\;\mathcal{S}^{RR}$ will be
associated with the lightest of the right-handed squarks, typically
$\tilde{t}_R$. In this framework, usually known as the light stop
scenario, to satisfy the out-of-equilibrium conditions the mass of the
right-handed stop should be small enough so that at the electroweak phase
transition temperature $T_\mathrm{ew}$ one has $
m_{\widetilde{t}_{R}}(T_\mathrm{ew}) \sim m_t(T_\mathrm{ew}) = h_t
v_{2}(T_\mathrm{ew})$. This is precisely the case we shall consider from
now on. In this context, the right-handed soft breaking masses are chosen
so that $m_{R1}^2 \simeq m_{R2}^2 \simeq m_{Q}^2, m_{R3}^2 \equiv m_{R}^2
\ll m_Q^2$. In particular, the parameter $m_R^2$ could be negative
provided that no charge or color breaking minima appear
\cite{Carena:1996wj}.

In this case,
\begin{equation}\label{2:CPSource}
\operatorname{Tr}\;\mathcal{S}^{RR}\simeq v^{2}(X)\;
\partial_{X}\beta(X)\; I_{RR}^{t}\operatorname{Im}\;\left(
\mu \left[Z_{R}\;{(Y_{u}^{A})}^{T}\;h_{u}^{*}\;
Z_{R}^{\dagger}\right]_{33}\right)\ .
\end{equation}
In the limit where the trilinear terms are flavor conserving, one can
easily recognize the well-known expressions obtained in Refs.
\cite{Riotto:1997vy,Carena:1997gx}.

\section{Electroweak baryogenesis}
\label{ebau}

In the present scenario, the baryon asymmetry generated in the broken
phase is determined by the density of left-handed quarks\footnote{In
principle, the baryon asymmetry is determined by the density of all
left-handed fermions, including leptons. However, in the usual electroweak
baryogenesis scenario, there is essentially no lepton asymmetry.},
$n_{L}$, created in front of the bubble wall in the symmetric phase
\cite{Huet:1995sh,Riotto:1997vy,Carena:1997gx,Cline:2000kb}. Such
densities induce weak sphalerons to produce a nonvanishing baryon number
\cite{'tHooft:up,Manton:1983nd,Kuzmin:1985mm}. If the system is near
thermal equilibrium and the particles are weakly interacting, the particle
densities $n_{i}$ are given by $n_{i}=k_{i}\mu_{i}T^{2}/6,$\ where
$\mu_{i}$ are the local chemical potentials and $k_{i}$ are statistical
factors equal to 2 (1) for bosons (fermions) and exponentially suppressed
for particles with masses $m_{i}\gg T.$ Assuming the supergauge
interactions to be in thermal equilibrium and neglecting all Yukawa
couplings except those corresponding to the top quark, it is possible to
express $n_{L}$ in terms of the densities of the chiral supermultiplet
$Q_{i}\equiv(q,\widetilde{q})\ $, $n_{Q_i}=n_{q_i} +n_{\widetilde{q}_i}$,
and of the right-handed top quark and squark, $T=t_{R}+\widetilde{t}_{R}$.
Indeed, assuming that all the quarks have nearly the same diffusion
constant, one obtains due to the strong sphaleron processes,
$n_{Q_{1}}=n_{Q_{2}}=2(n_{Q}+n_{T})$, implying the relation
\begin{equation}
n_{L}=n_{Q_{1}}+n_{Q_{2}}+n_{Q_{3}}=5n_{Q}+4n_{T}\ . \label{n_L}
\end{equation}

We can write down a set of coupled diffusion equations for the relevant
densities $n_{Q}\ $, $n_{T}\ $, $n_{H}=n_{H_{1}}+n_{H_{2}}\ $ and
$n_{h}=n_{H_{2}}-n_{H_{1}}\ $. Ignoring the curvature of the bubble wall,
all the quantities become functions of $z \equiv r+v_{w}t$, the coordinate
normal to the bubble wall surface. In the rest frame of the bubble wall we
obtain
\begin{align}
D_{q}n_{Q}^{\prime\prime}-v_{w}n_{Q}^{\prime}-\Gamma_{y}N_{1}-6\Gamma
_{ss}N_{2}-\Gamma_{m}N_{3}+\gamma_{Q}  &  =0,\label{eq_nQ}\\
D_{q}n_{T}^{\prime\prime}-v_{w}n_{T}^{\prime}+\Gamma_{y}N_{1}+3\Gamma
_{ss}N_{2}+\Gamma_{m}N_{3}-\gamma_{Q}  &  =0,\label{eq_nT}\\
D_{h}n_{H}^{\prime\prime}-v_{w}n_{H}^{\prime}+\Gamma_{y}N_{1}-\Gamma
_{h}\frac{n_{H}}{k_{H}}  &  =0,\label{eq_nH}\\
D_{h}n_{h}^{\prime\prime}-v_{w}n_{h}^{\prime}+\rho\Gamma_{y}N_{4}%
-(\Gamma_{h}+4\Gamma_{\mu})\frac{n_{h}}{k_{H}}  &  =0,\label{eq_nh}%
\end{align}
where, according to Eq.~(\ref{2:CPSource}), the $CP$-violating squark
current is given by
\begin{equation}\label{gammaQ}
\gamma_Q \simeq v_w\;v^{2}(z)\; \beta'(z)\;
I_{RR}^{t}\operatorname{Im}\;\left( \mu
\left[Z_{R}\;{(Y_{u}^{A})}^{T}\;h_{u}^{*}\;
Z_{R}^{\dagger}\right]_{33}\right)\ .
\end{equation}
Moreover,
\begin{align}
& N_{1} =\frac{n_{Q}}{k_{Q}}-\frac{n_{T}}{k_{T}}-\frac{n_{H}+\rho n_{h}
}{k_{H}}\ ,  & N_{2}=2\frac{n_{Q}}{k_{Q}}-\frac{n_{T}}{k_{T}}+9 \frac
{n_{Q}+n_{T}}{k_{B}}\ ,\nonumber\\
& N_{3} =\frac{n_{Q}}{k_{Q}}-\frac{n_{T}}{k_{T}}\ , & N_{4}=\frac{n_{Q}
}{k_{Q}}-\frac{n_{T}}{k_{T}}-\frac{n_{H}+n_{h}/\rho}{k_{H}}\ ,
\end{align}
with the parameter $\rho$ varying in the range from 0 to 1. The
coefficients $k_B=3\;$, $k_Q=6\;$, $k_T=9\;$, $k_H=12\;$ are statistical
factors and the quantities
\begin{align} \label{rates}
& \Gamma_{y}\simeq\frac{27\zeta(3)^{2}}{2\pi^{4}}h_{t}^{2}\;\alpha_{s}\; T
\simeq 2.4 \times 10^{-2}\;T\ ,\nonumber\\
& \Gamma_{ss} =16\; \alpha_s^4\; T \simeq 3.3 \times 10^{-3}\; T\ , \nonumber\\
& \Gamma_m = \frac{v^2(T)}{21\; T}\; h_t^2 \sin^2 \beta \simeq 4.8 \times
10^{-2}\; T\ ,\\
& \Gamma_h = \frac{v^2(T)}{140\; T} \simeq 7.1 \times 10^{-3}\; T\ , \nonumber\\
& \Gamma_\mu \simeq 0.1\;T\ ,\nonumber
\end{align}
are reaction rates: $\Gamma_{y}$ corresponds to the SUSY trilinear scalar
interaction involving the third generation squarks and the Higgs $H_1$
plus all SUSY and soft breaking trilinear interactions arising from the
superpotential term $h_t H_2 Q T$, $\Gamma_{ss}$ is the strong sphaleron
rate, $\Gamma_h$ and $\Gamma_m$ arise from the Higgs and axial top number
violating processes, while $\Gamma_\mu$ corresponds to the Higgs bilinear
term. The numerical estimates of these rates are obtained assuming the top
quark Yukawa coupling $h_t=1$, the strong coupling constant $\alpha_s
=0.12$, $v(T) \simeq T$ and $\tan\beta =10$. Finally, the diffusion
coefficients are given by
\begin{align} \label{diffcoeff}
D_q \simeq \frac{6}{T}\ , \quad D_h \simeq \frac{110}{T}\ .
\end{align}

Let us assume that $\Gamma_{y}$ and $\Gamma_{ss}$ are fast enough so that
$\Gamma _{y}D_{h}\ ,\;\Gamma_{ss}D_{q}\ ,\;\Gamma_{y}D_{q} \gg v_{w}^{2}.$
In this case
$N_{1}=\mathcal{O}(\Gamma_{y}^{-1}),\;N_{2}=\mathcal{O}(\Gamma_{ss}^{-1})$
and
\begin{equation}
\frac{n_{Q}}{k_{Q}}-\frac{n_{T}}{k_{T}}-\frac{n_{H}+\rho n_{h}}{k_{H}}%
\simeq0,\qquad2\frac{n_{Q}}{k_{Q}}-\frac{n_{T}}{k_{T}}+9\frac{n_{Q}+n_{T}%
}{k_{B}}\simeq0.
\end{equation}

This implies
\begin{align}
n_{Q}  &=
\frac{k_{Q}(9k_{T}-k_{B})}{k_{H}(k_{B}+9k_{Q}+9k_{T)}}(n_{H}+\rho
n_{h})\ ,\label{nQ}\\
n_{T}  &
=-\frac{k_{T}(9k_{T}+2k_{B})}{k_{H}(k_{B}+9k_{Q}+9k_{T)}}(n_{H}+\rho
n_{h})\ ,\label{nT}
\end{align}
and
\begin{equation}
N_{3}\simeq\frac{n_{H}+\rho n_{h}}{k_{H}},\qquad N_{4}\simeq N_{1}%
-\frac{1-\rho^{2}}{\rho}\frac{n_{h}}{k_{H}}\ .
\end{equation}
Thus, Eq. (\ref{eq_nh}) can be rewritten as:
\begin{equation}
D_{h}n_{h}^{\prime\prime}-v_{w}n_{h}^{\prime}+\rho\Gamma_{y}N_{1}%
-\left[(1-\rho^{2})\Gamma_{y}+\Gamma_{h}+4\Gamma_{\mu}\right]\frac{n_{h}}{k_{H}}=0.
\end{equation}

First we notice that for typical values of the scattering rate due to the
top quark Yukawa coupling \cite{Huet:1995sh}, the interaction rate
$(1-\rho^{2})\Gamma_{y}$\ is fast enough in a wide range of values of
$\rho$ (except those values very close to 1), leading to the solution
\begin{equation}
n_{h} \simeq 0.
\end{equation}
Indeed, assuming $v_{w}\simeq0.1,$ the condition
$(1-\rho^{2})\Gamma_{y}D_{h}\gtrsim v_{w}^{2}$ together with
Eqs.~(\ref{rates})-(\ref{diffcoeff}) imply that $\rho\lesssim0.998.$

Next, from Eqs. (\ref{nQ}), (\ref{nT}) and (\ref{n_L}) we find:
\begin{align}
n_{H}  &  \simeq\frac{k_{H}(k_{B}+9k_{Q}+9k_{T)}}{k_{Q}(9k_{T}-k_{B})}\;
n_{Q},\qquad\label{nHsol}\\
n_{T}  &  \simeq-\frac{k_{T}}{k_{Q}}\frac{9k_{Q}+2k_{B}}{9k_{T}-k_{B}}\;
n_{Q},\label{nTsol}\\
n_{L}  &  \simeq\frac{9k_{Q}k_{T}-5k_{Q}k_{B}-8k_{T}k_{B}}{k_{Q}(9k_{T}
-k_{B})}\;n_{Q}.\label{nLsol}%
\end{align}

Taking a linear combination of Eqs.~(\ref{eq_nQ})-(\ref{eq_nH}), which is
independent of the fast rates $\Gamma_{y}$ and $\Gamma_{ss}$ we obtain an
effective diffusion equation for $n_{Q}$ :
\begin{equation} \label{eq:nQfinal}
Dn_{Q}^{\prime\prime}-v_{w}n_{Q}^{\prime}-\Gamma n_{Q}+\gamma=0,
\end{equation}
where
\begin{align}
D  &  =\frac{1}{\Delta} \left[ D_{q}(9k_{Q}k_{T}+k_{Q}k_{B}+4k_{T}%
k_{B})+D_{h}k_{H}(k_{B}+9k_{Q}+9k_{T}) \right]  ,\\
\Gamma &  =\frac{k_{B}+9k_{Q}+9k_{T}}{\Delta}\left(  \Gamma_{m}+\Gamma
_{h}\right)  \theta(z),\\
\gamma &  =\frac{k_{Q}(9k_{T}-k_{B})}{\Delta}\;\gamma_{Q}\;\theta(z),\\
\Delta &  =9k_{Q}k_{T}+k_{Q}k_{B}+4k_{T}k_{B}+k_{H}(k_{B}+9k_{Q}+9k_{T})\
,
\end{align}
and $\theta(z)$ is the step function, which accounts for the fact that the
rates $\Gamma_m$ and $\Gamma_h$ are active only in the broken phase.

To estimate the baryon asymmetry we need to solve Eq.~(\ref{eq:nQfinal})
in the symmetric phase ($z\leq0$). Imposing the boundary conditions
$n_{Q}(\pm \infty)=0$ and the continuity of $n_{Q}$ and $n_{Q}^{\prime}$
at $z=0,$ a simple analytical approximation can be given:
\begin{equation} \label{nQanalytic}
n_{Q}(z)=A_{Q}e^{v_{w}z/D}\ ,
\end{equation}
where
\begin{equation}
A_{Q}=\frac{1}{\lambda_{+}D}\operatornamewithlimits{\int}_{0}^{\;\;\infty}
dz^{\prime} e^{-\lambda_{+}z^{\prime}}\gamma(z^{\prime}),\quad
\lambda_{\pm}=\frac{v_{w} \pm \sqrt{v_{w}^{2}+4\Gamma D}}{2D}\ .
\end{equation}

Assuming that the weak sphalerons are inactive in the broken phase (i.e.
inside the bubble), $n_{B}$ will be constant in this phase. To find the
value of this constant, one has to solve the diffusion equation for the
baryon asymmetry in the symmetric phase:
\begin{equation}
v_{w}n_{B}^{\prime}=-\theta(-z)\left[n_{F}\Gamma_{ws}n_{L}(z)+R\; n_{B}
(z)\right]\ ,
\end{equation}
where $n_{F}=3$ is the number of families, $R =
\frac{5}{4}n_{F}\Gamma_{ws}$ is the relaxation coefficient
\cite{Shaposhnikov:tw} and $\Gamma_{ws} \simeq 120\; \alpha_w^5 T$ is the
weak sphaleron rate. The above equation is easily integrated and we find
\begin{equation}
n_{B}=\frac{-n_{F}\Gamma_{ws}}{v_{w}}
\operatornamewithlimits{\int}_{-\infty}^{\;\;0}dz\;
e^{zR/v_{w}}\;n_{L}(z)\ .
\end{equation}
Using Eqs.~(\ref{nLsol}) and (\ref{nQanalytic}), one gets
\begin{equation} \label{nBfinal}
n_{B}=-n_{F}\Gamma_{ws}\frac{9k_{Q}k_{B}-8k_{B}k_{T}-5k_{B}k_{Q}}{k_{Q}
(9k_{T}-9k_{B)}}\left(  \frac{DA_{Q}}{DR+v_{w}^{2}}\right)  .
\end{equation}

To obtain a reliable approximation for the squark current, it is necessary
to specify the Higgs profiles as functions of $z$, i.e. the functions
$v(z)$ and $\beta(z)$ which appear in Eq.~(\ref{gammaQ}). A simple
expression is given by the kink approximation \cite{Carena:1997gx}
\begin{align}
v(z)  & =\frac{1}{2}v(T)\left[  1-\tanh\left( \alpha-\frac{2\alpha}{L_{w}
} z \right)  \right]\ ,  \label{Hprofile}\\
\beta(z)  & =\beta-\frac{1}{2}\Delta\beta\left[  1+\tanh\left( \alpha
-\frac{2\alpha}{L_{w}} z\right)  \right]\ ,  \label{betaprofile}
\end{align}
where $L_{w}/(2\alpha)$ parametrizes the thickness of the bubble wall,
$\alpha \simeq 3/2$ and $\Delta\beta$ is the variation of the angle
$\beta(z)$ along the bubble wall, which lies in the range $10^{-2} \gtrsim
\Delta\beta \gtrsim 4 \times 10^{-3}$ for $100$~GeV $\leq m_A \leq
200$~GeV \cite{Moreno:1998bq}. At first order in $v_{w}$, all the
dependence in $z$ of the source $\gamma_{Q}$ is given by the combination $
v^{2}(z)\beta^{\prime}(z)\ $. Substituting the profiles (\ref{Hprofile})
and (\ref{betaprofile}) into $A_{Q}$ (cf. Eq.~(\ref{nQanalytic})), it is
possible to integrate explicitly the $z$ dependence. Finally we have
\begin{equation} \label{AQfinal}
A_{Q}=\frac{v^{2}(T)\Delta\beta}{48D\lambda_{+}}\;\frac{k_{Q}(9k_{T}
-k_{B})}{\Delta}\;v_{w}\operatorname{Im}\,\left(  \mu\left[
Z_{R}\;{(Y_{u}^{A})}^{T}\;h_{u}^{\ast}\;Z_{R}^{\dagger}\right]
_{33}\right) I_{RR}^{t}\; f_{\lambda_{+}},
\end{equation}
with $I_{RR}^{t}$ given by Eqs.~(\ref{2:IRR})-(\ref{2:IpGp}),
\begin{align} \label{fplus}
f_{\lambda_{+}}  = & \frac{2a(a-4)(a-2)}{a-6}\; e^{-6 \alpha}\;
{}_{2}F_{1}(1,3-\frac{a}{2};4- \frac{a}{2};-e^{-2\alpha })
+a(a-4)(a-2)\;\pi\; e^{-a \alpha} \operatorname{cosec} (a \pi/2)
\nonumber\\
& -2(1+e^{2\alpha })^{-3}\;e^{-4 \alpha}\left[ (a^{2}+8)e^{4\alpha}
+2a(a-1)e^{2\alpha}+a(a-2)\right]  ,
\end{align}
$a = \lambda_{+}L_{w}/(2\alpha)$ and $_{2}F_{1}(a,b;c;z)$ is the
hypergeometric function.

In Section~\ref{numerical} we will present our numerical results for the
baryon asymmetry using the approximate solution given by
Eqs.~(\ref{nBfinal}), (\ref{AQfinal}) and (\ref{fplus}).

\section{Baryon asymmetry in supersymmetric models}
\label{textures}

In this section we study the flavor-dependent contribution to the
squark current, which enters in Eq.~(\ref{2:CPSource}) through the
imaginary part
\begin{equation} \label{SRRpropto}
\operatorname{Tr}\;\mathcal{S}_{RR} \propto \operatorname{Im}\,\left(
\mu\left[ Z_{R}\;{(Y_{u}^{A})}^{T}\;
h_{u}^{*}\;Z_{R}^{\dagger}\right]_{33} \right)\ .
\end{equation}
Our goal is to maximize the above quantity and, simultaneously, satisfy
the experimental constraints on the off-diagonal $A$ terms as well as on
the CKM mixing matrix.

To establish a direct connection between $\mathcal{S}_{RR}$ and low energy
supersymmetry phenomenology, let us express $\mathcal{S}_{RR}$ in terms of
the up left-right mass insertions, $\delta_{LR}^{u}$:
\begin{equation}\label{deltaLRdef}
\left(  \delta_{LR}^{u}\right)  _{ij}\equiv\frac{v_{2}\left(  U^{L}\left(
Y_{u}^{A}\right)  ^{*}U^{R\dagger}\right)_{ij}-v_{1}
\mu\left(d_{u}\right)_{ij}}{\langle m_{\widetilde{q}}^2\rangle}\ ,
\end{equation}
normalized to $\left\langle m_{\widetilde{q}}^2\right\rangle $, the mean
value of the squark masses; $U^{L,R}$ are the unitary matrices that
diagonalize $h_u$,
\begin{equation}
U^{L}h_{u}^{*}U^{R\dagger}=d_u =
\frac{1}{v_2}\operatorname{diag}\;(m_u,m_c,m_t)\ .
\end{equation}

For the sake of simplicity and to maximize
$\operatorname{Tr}\;\mathcal{S}_{RR}\ $, let us assume that
$Z_{R}=\openone$. We notice that the unitarity of $Z_R$ implies that any
deviation from the identity (or its permutations) will introduce
additional suppression factors in the $CP$-violating source. From
Eqs.~(\ref{SRRpropto}) and (\ref{deltaLRdef}) we obtain
\begin{equation}
\operatorname{Tr}\;\mathcal{S}_{RR}\propto \mu h_{t}^{2}\frac{\left\langle
m_{\widetilde{q}}^2\right\rangle}{m_{t}}\operatorname{Im}\,\left[
U_{i3}^{R*}U_{33}^{R}\left(  \delta_{LR}^{u}\right)_{3i}^{*}\right]\ .
\end{equation}

\begin{figure}[ht]
\begin{center}
\includegraphics[width=10cm]{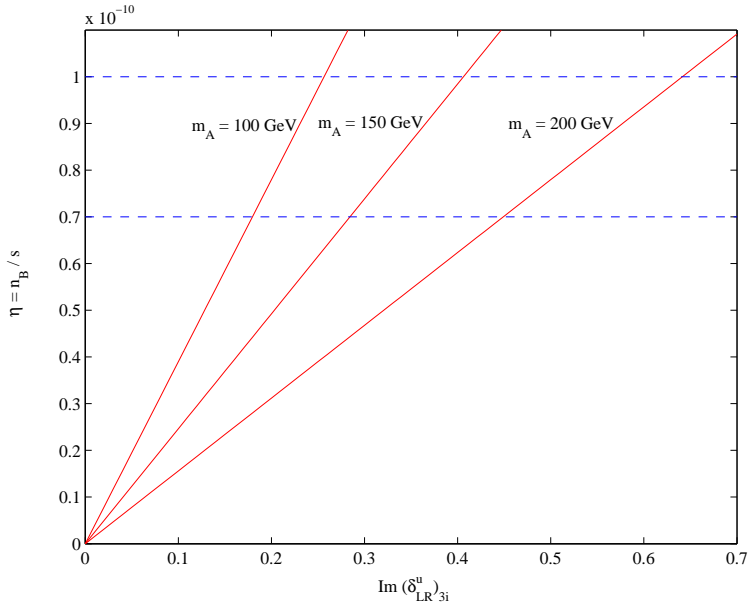} \\
\caption{Baryon asymmetry as a function of $\operatorname{Im}\;
(\delta_{LR}^u)_{3i}$ for different values of $m_A$.  We assume
$\mu=500$~GeV, $v_w=0.04$ and $\langle m_{\widetilde{q}}^2\rangle \simeq
m_{Q}^2 = 1$~TeV$^2$. The other parameters are indicated in
Table~\ref{table1}. The dashed lines correspond to the lower and upper
bounds of the observed baryon asymmetry given in Eq.~(\ref{nBsbound}). }
\label{fig2}
\end{center}
\end{figure}

\begin{figure}[ht]
\begin{center}
\includegraphics[width=10cm]{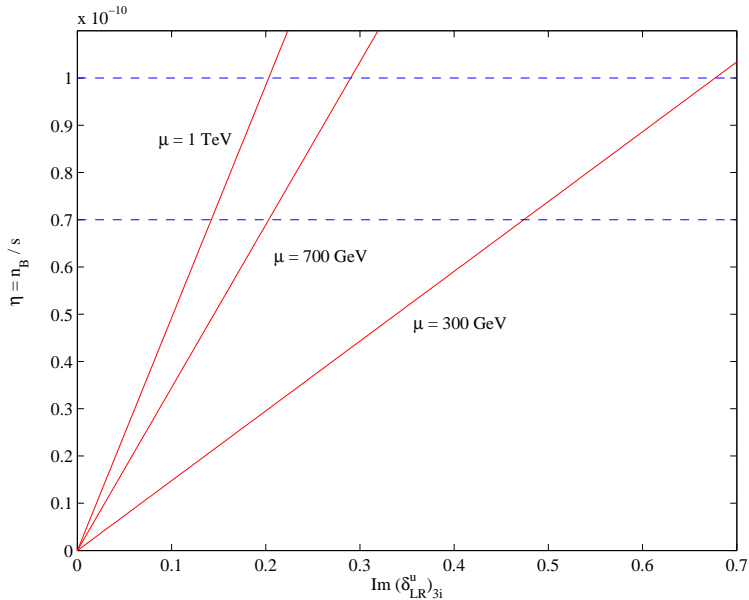} \\
\caption{Baryon asymmetry as a function of $\operatorname{Im}\;
(\delta_{LR}^u)_{3i}$ for different values of $\mu$.  We assume
$m_A=150$~GeV, $v_w=0.04$ and $\langle m_{\widetilde{q}}^2\rangle \simeq
m_{Q}^2 = 1$~TeV$^2$. The other parameters are taken as in
Table~\ref{table1}. The dashed lines correspond to the lower and upper
bounds of the observed baryon asymmetry given in Eq.~(\ref{nBsbound}). }
\label{fig3}
\end{center}
\end{figure}

Assuming for instance that the dominant contribution comes from large
mixing between the $t$- and either the $u$- or $c$-quark, i.e. taking
$U_{i3}^{R} = U_{33}^{R}=1/\sqrt{2}\ (i=1,2)$, we have
\begin{equation}
\operatorname{Tr}\;\mathcal{S}_{RR}\sim\frac{1}{2}\mu
h_{t}^{2}\frac{\left\langle
m_{\widetilde{q}}^2\right\rangle}{m_{t}}\operatorname{Im}\,\left[ \left(
\delta_{LR}^{u}\right)_{3i}^{*}\right]\ .
\end{equation}

In Fig.~\ref{fig2} we plot the baryon asymmetry as a function of
$\operatorname{Im}\; (\delta_{LR}^u)_{3i}$ for different values of $m_A$.
It is seen that even for light pseudo-scalar masses $m_A$, large values of
$\operatorname{Im}\; (\delta_{LR}^u)_{3i}$ are required in order to
reproduce the observed BAU. In Fig.~\ref{fig3} the baryon asymmetry is
given as a function of $\operatorname{Im}\; (\delta_{LR}^u)_{3i}$ for
different values of $\mu$. We notice that it is possible to get a
sufficient BAU if
\begin{align} \label{deltaLRlowbound}
\operatorname{Im}\,\left[ \left( \delta_{LR}^{u}\right)_{3i}^{*}\right]
\gtrsim 0.14\ , \quad \text{for} \quad \mu \simeq 1~\text{TeV}\ .
\end{align}

It is worth noticing that these bounds on the imaginary part of the mass
insertions $(\delta^u_{LR})_{3i}$, $i=1,2$ are compatible with the bounds
obtained from the chargino contributions to $B_d - \bar{B}_d$ mixing and
the $CP$ asymmetry of $B_d \to J/\psi K_S$ \cite{Gabrielli:2002fr}. In
addition to fulfilling the  bounds on the $LR$ mass insertions that arise
from satisfying the BAU requirements (as shown in Figs.~\ref{fig2} and
\ref{fig3}), the textures for the Yukawa couplings and trilinear terms
must be compatible with the FCNC constraints \cite{Gabbiani:1996hi},
charge and color breaking bounds \cite{Casas:1996de} and those associated
with the measurements of the electric dipole moments of the neutron and
mercury atom. Regarding the latter, the current experimental bounds
are~\cite{Harris:jx}
\begin{align}
d_n < 6.3 \; \times 10^{-26} \; e\; \text{cm}\;\ , \quad \quad d_{Hg} <
2.1 \; \times 10^{-28} \; e\; \text{cm}\ .
\end{align}
These values can be translated into bounds for the imaginary parts of the
up and down $(\delta_{LR})_{11,22}$~\cite{Abel:2001vy}. In particular,
\begin{align}
\operatorname{Im}\; (\delta_{11}^u)_{LR} <\;10^{-6} \quad \text{and}\quad
\operatorname{Im}\; (\delta_{11}^u)_{LR} <\;10^{-7} - 10^{-8}\ ,
\end{align}
from the neutron and mercury atom EDM's, respectively.

Besides the underlying EDM problem that is associated with the imaginary
parts of the diagonal elements of $\delta^u_{LR}\,$, one should also bear
in mind the extensive array of constraints on the non-diagonal entries.
These stem from FCNC bounds, limits on rare decays and the observed amount
of $CP$-violation in the $K$ and $B_d$ meson systems~\cite{Buras:1997ij}.

We can also express Eq.~(\ref{SRRpropto}) in a different manner using the
definition (\ref{2:YAdef}). Assuming $\mu$ real and keeping only terms
proportional to $h_{t}$, one has
\begin{equation}
\operatorname{Tr}\;\mathcal{S}_{RR} \propto \mu h_{t}^{2}\left|
U_{3k}^{L}\right| ^{2}\operatorname{Im}\left[ \left(  Z_{R}\right)
_{3j}A_{kj}^{u}U_{3j}^{R*}U_{3l}^{R}(Z_{R}^{*})_{3l}\right]\ ,
\end{equation}
where a summation over the indices $j,k,l=1,2,3$ is understood. In this
case, the imaginary part of $\operatorname{Tr}\;\mathcal{S}_{RR}$ is given
by the simple form:
\begin{equation}
\operatorname{Tr}\;\mathcal{S}_{RR} \propto \mu h_{t}^{2}\left|
U_{3k}^{L}\right|^{2}\left|  U_{33}
^{R}\right|^{2}\operatorname{Im}\left(A_{k3}^{u}\right)\ ,
\end{equation}
where $Z_R = \openone$ has been assumed. The important point is that
$\mathcal{S}_{RR}$ is proportional to the imaginary part of
$A_{k3}^{u}\,$, which are quantities weakly constrained by experimental
data in contrast to other terms proportional to $A_{ij}$ ($i,j=1,2$)
\cite{Gabbiani:1996hi}.

In the case of hierarchical Yukawa couplings, the matrix $U^{R}$ can be
assumed to be very close to the identity and the maximum values for
$U^{L}$ are bounded by the CKM matrix. In such scenario, we have
\begin{align}
\operatorname{Tr}\;\mathcal{S}_{RR} \propto \mu h_{t}^{2}\left|
U_{3k}^{L}\right| ^{2}\operatorname{Im} \left(  A_{k3}^{u}\right) \lesssim
\mu h_{t}^{2}\left| V_{3k}^{CKM}\right|^{2}\operatorname{Im} \left(
A_{k3}^{u}\right)\ .
\end{align}
It is easy to see that the contribution proportional to $A_{13}^{u}$ and
$A_{23}^{u}$ is very suppressed by $\left|V_{td}\right|^{2}$ and $\left|
V_{ts}\right|^{2}$ and thus, the dominant contribution comes from the
imaginary part of $A_{33}^{u}$ which is always constrained in GUT-inspired
SUSY models by the EDM bounds. In particular, in the minimal supergravity
inspired model where the $A$-terms are universal and the overall phase is
constrained to be $\lesssim 10^{-1}$, this contribution becomes quite
negligible and one cannot obtain a viable BAU through the squark current.

Therefore, the flavor-dependent phases will play an important role in
baryogenesis only if significant mixing between the top quark and up
and/or charm quark is present. This mixing can be obtained in
supersymmetric models with non-universal soft breaking terms and, for
instance, with nearly democratic Yukawa couplings. Of course, this means
that processes like $t\rightarrow u+\gamma,\; t \rightarrow c+\gamma$ may
have much larger branching ratios than those predicted in the SM.

To illustrate our results, let us now consider some simple textures for
the quark and squark mass matrices at the electroweak scale. In case A we
present the generic case of a SUSY texture where the CKM matrix is
dominated by the down quark mixing. Case B corresponds to SUSY models with
Hermitian Yukawa couplings and trilinear terms at GUT scale. Of course,
the RGE running down to the electroweak scale will induce some deviations
from hermiticity, but their contributions will not be relevant to our BAU
analysis. In case C, the textures are chosen such that the BAU produced at
the electroweak scale is maximized and the phenomenological constraints
are still satisfied.

\bigskip

\noindent \textbf{Case A}

We first consider the following simple texture for the trilinear matrix
$A_u$:
\begin{equation}
\label{caseA:Au} A_u= A_t \left(
\begin{array}[c]{ccc}
1\; & 1 & 1\\
1\; & 1 & 1\\
1\; & 1 & a\;e^{i\pi/2}
\end{array}
\right)\ ,
\end{equation}
with $a=\mathcal{O} (1)$ a real parameter. We also assume that the up
quark Yukawa coupling matrix $h_u$ is diagonal, i.e. $h_u = d_u,\; U_L =
\openone\ ,\; U_R = \openone$. In this case one obtains
\begin{equation}
\operatorname{Tr}\;\mathcal{S}_{RR} \propto \mu
h_{t}^{2}\operatorname{Im}\left( A_{33}^{u}\right) = a A_t \mu h_{t}^{2}\
.
\end{equation}

\bigskip

\noindent \textbf{Case B}

Let us now assume that the up- and down-quark Yukawa couplings as well as
the trilinear ones are Hermitian matrices. In this case the up-quark
Yukawa coupling matrix is diagonalized by a unitary matrix $U$ such that
$h_u=U^{T} d_u U^*$. Since the trilinear matrix $A_u$ is Hermitian too,
then $\operatorname{Im}\;A_{33}=0$. The squark source $\mathcal{S}_{RR}$
can be expressed as
\begin{align}
\operatorname{Tr}\;\mathcal{S}_{RR} \sim\mu h_{t}^{2}\left|
U_{33}\right|^{2}\left\{ \left| U_{31} \right|^{2}\operatorname{Im}\left(
A_{13}^{u}\right) +\left| U_{32}\right|^{2}\operatorname{Im}\left(
A_{23}^{u}\right)  \right\}\ .
\end{align}

It is easy to see that in order to maximize the imaginary part of
$\mathcal{S}_{RR}$, a large mixing between the third and the first (or the
second) families of up quarks is needed. For instance, assuming maximal
mixing between the $c$ and $t$ quarks, i.e.
\begin{equation}
U= \left(
\begin{array}[c]{ccc}
1\; & 0 & 0\\
0\; & 1/\sqrt{2} & -1/\sqrt{2}\\
0\; & 1/\sqrt{2} & 1/\sqrt{2}
\end{array}
\right)\ ,
\end{equation}
and the following Hermitian texture for the trilinear matrix $A_u$,
\begin{equation}
\label{caseB:Au} A_u= A_t \left(
\begin{array}[c]{ccc}
1\; & 1 & 1\\
1\; & 1 & b\;e^{i\pi/2}\\
1\; & b\;e^{-i\pi/2} & 1
\end{array}
\right)\ ,
\end{equation}
with $b$ a real parameter of order $\mathcal{O}(1)$, one obtains
\begin{equation}
\operatorname{Tr}\;\mathcal{S}_{RR} \propto \frac{1}{4}\mu
h_{t}^{2}\operatorname{Im}\left( A_{23}^{u}\right) = \frac{b}{4} A_t \mu
h_{t}^{2}\ .
\end{equation}
The corresponding texture for the down-quark Yukawa matrix $h_d$ is fixed
by the structure of the CKM matrix, $V_{CKM}= UD^{\dagger}\;$, where $D$
is the mixing matrix that diagonalizes the down quark mass matrix,
$h_d=D^T d_d D^*$, with
$d_d=\dfrac{1}{v_1}\operatorname{diag}\;(m_d,m_s,m_b)$. Thus, we have
\begin{equation}
h_{d}=U^{T}V_{CKM}^{*} d_d V_{CKM}^{T} U^\ast\ .
\end{equation}

\bigskip

\noindent \textbf{Case C}

In this case we take
\begin{equation}
\label{caseC:Au} A_u= A_t \left(
\begin{array}[c]{ccc}
1\; & 1 & c\;e^{i\pi/2}\\
1\; & 1 & 1\\
1\; & 1 & 1
\end{array}
\right)\ , \quad
 U_L= \left(
\begin{array}[c]{ccc}
0\; & 0\; & 1\\
0\; & 1\; & 0\\
1\; &0\; & 0
\end{array}
\right)\ , \quad U_R = \openone\ ,
\end{equation}
where $c=\mathcal{O}(1)$. The squark source $\mathcal{S}_{RR}$ is then
given by
\begin{equation}
\operatorname{Tr}\;\mathcal{S}_{RR} \propto \mu h_{t}^{2}
\operatorname{Im}\left( A_{13}^{u}\right) = c A_t \mu h_{t}^{2}\ .
\end{equation}

Let us remark that in the previous examples most of the entries in the
trilinear matrix $A_u$ were set for simplicity equal to one. It is clear
that in general those elements can be nondegenerate and even complex.

In the next section we shall present some numerical examples for the
textures considered above and compute the BAU generated by the
corresponding $CP$-violating squark current.

\section{Numerical results}
\label{numerical}

In order to produce the required BAU, it is needed not only to violate the
$CP$ and $C$ symmetries, but the universe has also to be out of
equilibrium during the stage of baryogenesis, to avoid any washout by the
electroweak sphalerons. This can happen if the electroweak phase
transition is of first-order. To freeze the action of the sphalerons,
their rate in the broken phase has to be smaller than the Hubble rate,
i.e.
\begin{equation}
\Gamma_\mathrm{ew}(T_\mathrm{c})<H(T_\mathrm{c})\ , \nonumber
\end{equation}
where $T_\mathrm{c}$ is the critical temperature of the phase transition.
Using the sphaleron rate in the broken phase \cite{Arnold:1987mh}, this
equation is translated into a condition on the vacuum expectation value of
the Higgs fields at $T_\mathrm{c}$:
\begin{equation} \label{vToverT}
\frac{v(T_\mathrm{c})}{T_\mathrm{c}}\gtrsim 1\ .
\end{equation}

In the standard model of electroweak interactions, this would imply a
Higgs mass lighter than 43 GeV, which is already experimentally ruled
out~\cite{Hagiwara:pw}. A way to avoid this constraint is to add scalar
fields (with Higgs field-dependent masses) such that their contribution to
the finite temperature potential will increase the strength of the phase
transition. Naturally, the MSSM is an appealing alternative to the SM. In
particular, it has been shown that the presence of a light right-handed
top squark with small mixing considerably enhances the strength of the
phase transition~\cite{Carena:1996wj,Espinosa:1996qw,Delepine:1996vn}.
However, in this case a few constraints need to be satisfied. Within the
MSSM, all these requirements impose restrictions to the allowed parameter
space \cite{Quiros:2000wk}: (i) a heavy pseudoscalar mass $m_A$ and a
large $\tan \beta$ regime, $m_{A} \gtrsim 150$~GeV, $\tan \beta \gtrsim
5$; (ii) heavy left-handed stops with a relatively small mixing, $m_{Q}
\gtrsim 1$~TeV, $0.25 \lesssim (A_{t}-\mu \cot \beta)/m_{Q} \lesssim 0.4$;
(iii) a light right-handed stop, $105$~GeV $\lesssim m_{\widetilde{t}_{R}}
\lesssim 165$~GeV.

In the context of models with non-universal $A$ terms, we can expect to
have more freedom and that some of these conditions may be relaxed. In
fact, if we want to keep alive the light stop scenario, the constraints
imposed by the electroweak phase transition (cf.~Eq.~(\ref{vToverT})) are
typically the same than in the one squark generation case, except that now
we should require
\begin{equation} \label{deltaLR:constraint}
|(\delta_{LR}^u)_{ij}|  \lesssim 0.4 \times \frac{m_t m_Q}{\langle
m_{\widetilde{q}}^2 \rangle} \approx 0.07\ ,
\end{equation}
if one assumes that $\langle m_{\widetilde{q}}^2 \rangle=m_Q^2=1$ TeV$^2$.

From the last equation and the bound given in Eq.~(\ref{deltaLRlowbound})
(see also Figs.~\ref{fig2} and \ref{fig3}), we conclude that it is
difficult to produce enough BAU using the flavor-dependent phases of the
trilinear soft breaking terms unless $\mu \gtrsim 1$~TeV.

\begin{table*}
\caption{Input parameters used in our numerical calculations
 \cite{Huet:1995sh,Riotto:1997vy,Carena:1997gx}.}
\begin{tabular*}{12cm}{lll}
\hline \hline
\multicolumn{2}{l}{\textsc{MSSM parameters}}\\
$m_W= 80.4$~GeV  & $m_Z=91.2$~GeV & $v=246.22$~GeV\\
$m_{u}=2.3$~MeV & $m_{c}= 0.68$~GeV & $m_{t}=174.3$~GeV\\
$A_t =400$~GeV & $m_Q = 1$~TeV & $m_R=0$\\
$m_{H}=115$~GeV &  $\tan \beta = 10$ & $100 \leq m_A \leq 200$~GeV\\
\multicolumn{2}{l}{\textsc{BAU parameters}}\\
$T= 100$~GeV & $g_\ast = 125.75$ &  \\
$L_w=20/T$ & $\alpha = 3/2$ & $4 \times 10^{-3} \lesssim
\Delta\beta \lesssim 10^{-2}$\\
\multicolumn{2}{l}{\textsc{Statistical factors}}\\
$k_B=3$ & $k_H=12$ & \\
$k_Q=6$ & $k_T=9$ & \\
\multicolumn{2}{l}{\textsc{Diffusion coefficients}}\\
$D_q=6/T$ & $D_h=110/T$ & \\
\multicolumn{2}{l}{\textsc{Reaction rates}}\\
$\Gamma_\mu=0.1 \; T$& $\Gamma_{ws}=5.4\times10^{-6}\;T$& \\
$\Gamma_{ss}=3.3\times10^{-3}\;T$ &$\Gamma_{y}=2.4\times10^{-2}\;T$ &\\
$\Gamma_m=4.8\times10^{-2}\;T$&$\Gamma_h=7.1\times10^{-3}\;T$& \\
$\Gamma_{Q}=\Gamma_{R}=\Gamma_{\tilde{t}}=0.12\; T$ &  &  \\
\hline
\end{tabular*}  \label{table1}
\end{table*}

\begin{figure}[ht]
\begin{center}
\includegraphics[width=10cm]{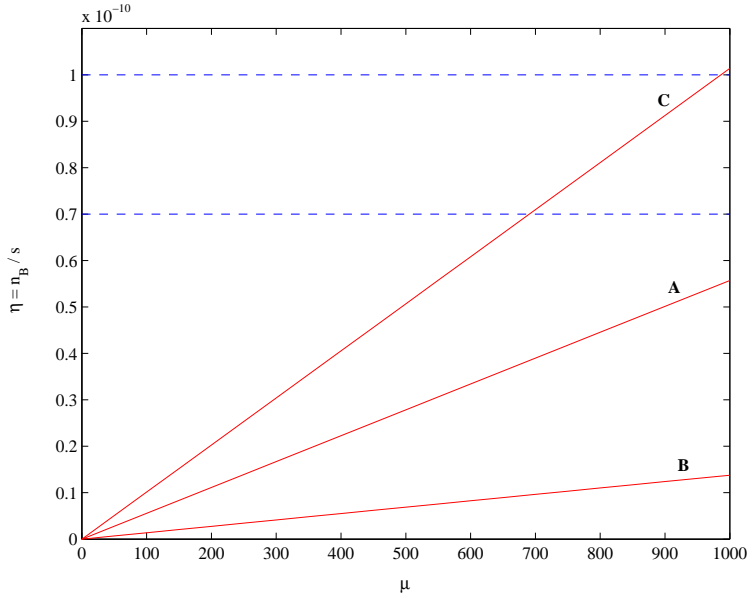}
\caption{The baryon-to-entropy ratio $n_B/s$ as a function of the $\mu$
parameter for different up-quark Yukawa coupling and trilinear matrix
textures (cases A, B and C considered in Section \ref{textures}). We have
chosen $a=1,\; b=1$ and $c=1.8$ for the coefficients in the textures
(\ref{caseA:Au}), (\ref{caseB:Au}) and (\ref{caseC:Au}), respectively. We
assume $m_A=150$~GeV and $v_w=0.04\,$. The rest of the parameters are
chosen according to Table~\ref{table1}. The dashed lines correspond to the
lower and upper bounds of the observed baryon asymmetry.} \label{fig4}
\end{center}
\end{figure}

In Fig.~\ref{fig4} we present the baryon-to-entropy ratio $n_B/s$ as a
function of the $\mu$ parameter for different up-quark Yukawa coupling and
trilinear term textures (cases A, B and C considered in Section
\ref{textures}). The coefficients $a,\;b,\;c$ that appear in the textures
(\ref{caseA:Au}), (\ref{caseB:Au}) and (\ref{caseC:Au}), are chosen so
that the $CP$-violating squark source is maximized and the constraints
coming from the electroweak phase transition are satisfied. We also
require the lightest Higgs mass to be consistent with the present
experimental lower bound, $m_H \gtrsim 109$~GeV~\cite{Hagiwara:pw}. In
particular, to maximize the $CP$-violating source and, simultaneously,
satisfy the bounds on the lightest right-handed squark mass,
$m_{\tilde{t}_R} \gtrsim 105$~GeV, we take $a=1,\; b=1$ and $c=1.8\ $.
From the figure it is seen that textures A and B cannot generate enough
BAU. This is related to the fact that the coefficients $a$ and $b$ cannot
be increased without violating the bounds on $m_{\tilde{t}_R}\ $. It is
also clear that the produced BAU in the Hermitian case (B) is further
suppressed by a factor 1/4 due to the Yukawa quark mixings. On the other
hand, texture C can produce the observed baryon asymmetry provided that
the $\mu$ parameter is large enough. We also notice that in this case the
coefficient $c$ is less constrained by $m_{\tilde{t}_R}\ $.

\begin{figure}[ht]
\begin{center}
\includegraphics[width=10cm]{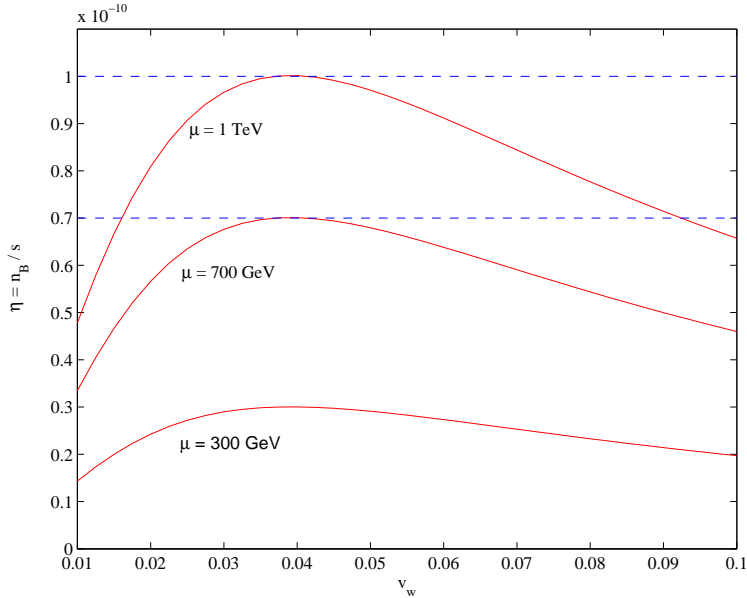}
\caption{The baryon-to-entropy ratio $n_B/s$ as a function of the wall
velocity $v_w$ for different values of $\mu$ and $m_A=150$~GeV. The other
parameters are chosen as in Table~\ref{table1}. We assume the textures
given in case C of Section~\ref{textures}.  The dashed lines correspond to
the lower and upper bounds of the observed baryon asymmetry given in
Eq.~(\ref{nBsbound}).} \label{fig5}
\end{center}
\end{figure}

The baryon asymmetry as a function of the wall velocity $v_w$ is plotted
in Fig.~\ref{fig5} for different values of $\mu$ and assuming the textures
of case C, which succeeds in producing enough BAU. As we can see from the
figure, to get a BAU compatible with the observational limits on $n_{B}/s$
(cf. Eq.~(\ref{nBsbound})) large values of $\mu$ ($\mu \gtrsim 700$~GeV)
are required.

We should emphasize however that the electroweak baryogenesis is a
strongly out-of-equilibrium process and all our computation is based on
several approximations. Unfortunately, it is very difficult to estimate
the errors done during the calculation because typically they come from
different sources. Nevertheless, we can conclude that the scenario with a
light stop and $CP$ violation coming from flavor-dependent phases is in
general disfavored. Of course, the constraints coming from the electroweak
phase transition can be in principle relaxed in extensions of the MSSM
with new scalar fields, for example by adding singlet fields (NMSSM)
\cite{Pietroni:in}.

In the present model, where the $CP$-violating sources arise from the
flavor-dependent phases of the SUSY soft breaking trilinear terms, the
strongest constraint comes from the lightest up squark mass. Indeed, two
conditions that will push the lightest up-squark mass to smaller values
must be fulfilled. First, $m_{R} \lesssim T_{\mathrm{ew}}$, otherwise the
contribution of the lightest right-handed squark to the BAU will be
exponentially suppressed by a Boltzmann factor. Secondly, as we have seen
from Fig.~\ref{fig3}, $\operatorname{Im} \left(\delta_{LR}^u\right)_{3i}
\gtrsim 0.14$, which typically means that some of the $A_{ij}$'s have to
be of the order or larger than $m_Q$. In other words, the $6 \times 6$
up-squark mass matrix will have large mixings.

In this paper, we have restricted ourselves to the range of parameters
satisfying the upper bound (\ref{deltaLR:constraint}) such that all the
experimental and theoretical constraints on the squarks and the lightest
Higgs  are satisfied in the light stop scenario. An important question
that remains is whether it is possible to find GUT patterns for the
trilinear terms, SUSY soft-breaking squark masses and quark Yukawa
couplings, such as to obtain large values of $\delta_{LR}^u$ and a strong
first-order phase transition, while still satisfying the experimental
limits on the squark and Higgs masses, as well as the EDM's constraints.

\section{Conclusions}
\label{conclusion}

Recent EDM bounds impose severe constraints on the usual scenario of
supersymmetric electroweak baryogenesis based on flavor-conserving
$CP$-violating phases. A natural solution to this problem is to work in
the framework where all the flavor-conserving parameters, such as the
$\mu$-term and gaugino masses are real. In this case, the dominant
contribution to the baryon asymmetry is associated to the flavor-dependent
$CP$-violating squark sources.

In this work we have studied in detail the impact of non-universal $A$
terms on the scenario of electroweak baryogenesis. By generalizing the
standard approach, we have obtained the expression for the $CP$-violating
squark sources with explicit flavor dependence. We have shown that if we
impose on these terms the condition to have a strong first-order phase
transition induced by a light right-handed squark, the baryon asymmetry of
the universe produced at the electroweak scale is typically too small,
thus disfavoring this scenario.

On the other hand, if we assume that the problem of the strength of the
first-order electroweak phase transition is solved through another
mechanism as it can happen in extensions of the MSSM with additional Higgs
scalars, it is possible to have textures for the up- and down-quark Yukawa
coupling matrices in order to maximize the $CP$-violating source. This
however implies a large mixing between the top quark and one of the light
up-quarks ($u$ or $c$) as well as large deviations from universality for
the $A$ terms (typically, $\left( \delta_{LR}^{u}\right)_{13,23} \gtrsim
0.14$). Such large $\delta_{LR}^{u}$ could have important implications on
flavor-changing top decays.

Supersymmetric electroweak baryogenesis is an attractive mechanism to
explain the observed baryon asymmetry of the universe. Not only the
physics involved in this process is directly related to low-energy
observables, but it can also be testable in accelerator experiments in the
near future. In particular, searches for a light Higgs boson and a light
stop at LHC and Tevatron will constitute a test of the viability of this
scenario.

There are still a few questions to be answered and controversial issues to
be clarified. For instance, the precise details of the electroweak phase
transition are still unknown and there is at present a debate regarding
the structure of the $CP$-violating currents that are relevant to
baryogenesis. Nevertheless, a considerable progress has been done during
the past few years, and the effort directed to resolve these problems
could give us definite answers in the near future.

\begin{acknowledgements}

We thank Marcos Seco for useful discussions and comments. The work of D.D.
was supported by \emph{Funda\c{c}\~{a}o para a Ci\^{e}ncia e a Tecnologia}
(FCT) through the project POCTI/36288/FIS/2000. The work of R.G.F. has
been supported by FCT under the grant SFRH/BPD/1549/2000. The work of S.K.
was supported by PPARC. A.T. acknowledges the support from FCT under the
grant PRAXIS XXI BD/11030/97.
\end{acknowledgements}

\end{document}